\documentclass[aps,twocolumn,showpacs]{revtex4}%
\usepackage{amsfonts}
\usepackage{amsmath}
\usepackage{amssymb}
\usepackage{appendix}
\usepackage{graphicx}%
\providecommand{\U}[1]{\protect\rule{.1in}{.1in}}
\providecommand{\U}[1]{\protect\rule{.1in}{.1in}}

\begin{document}
\title{Optomechanically induced transparency and gain}
\author{\ Xiao-Bo Yan}
\email{xiaoboyan@126.com}
\affiliation{College of Electronic Science, Northeast Petroleum University, Daqing  163318, China}
\date{\today}


\begin{abstract}
Optomechanically induced transparency is an important quantum phenomenon in cavity optomechanics. Here, we study the properties of optomechanically induced transparency in the simplest optomechanical system (consisting of one cavity and one mechanical resonator) considering the effect of non-rotating wave approximation (NRWA) that was ignored in previous works. With the NRWA effect, we find the ideal optomechanically induced transparency dip can be easily achieved, and the width of optomechanically induced transparency dip can become very narrow especially in unresolved sideband regime. Finally, we study the properties of optomechanically induced gain, and give the analytic expression about the maximum value of gain.
\end{abstract}
\maketitle

\section{Introduction}

Cavity optomechanics \cite{Aspelmeyer2014} exploring the interaction between
macroscopic mechanical resonators and light fields, has received increasing
attention for the broad applications in testing macroscopic quantum physics,
high-precision measurements, and quantum information processing
\cite{Aspelmeyer2014,Kippenberg2008,Marquardt2009,Verlot2010,Mahajan2013}.
Various experimental systems exhibiting such
interactions are proposed and investigated, such as Fabry-Perot cavities \cite{Gigan2006,Arcizet2006}, whispering-gallery microcavities \cite{Kippenberg2005,Tomes2009,Jiang2009}, membranes \cite{Thompson2008,Jayich2008,Sankey2010,Karuza2013}, and superconducting circuits \cite{Regal2008,Teufel2011_471}.
In these optomechanical systems, the motion of mechanical oscillator can be
effected by the radiation pressure of cavity field, and this interaction can
generate various quantum phenomena, such as ground-state cooling of mechanical
modes
\cite{Marquardt2007,Wilson-Rae2007,LiuYC2013,BingHe2017,Wang2018,Stefanatos2016}, quantum entanglement \cite{Liao2014,Deng2016,Vitali2007,Yan2017,Yan2019OE,Li2017,Stefanatos2017}, nonclassical mechanical states \cite{Nation2013,Ren2013,Bergholm2019,Meng2019}, normal mode splitting
\cite{Nature460,Dobrindt2008,Huang2009}, and nonreciprocal optical transmissions \cite{Xu2015,BingHe2018,Yan2019FOP}, etc.

Optomechanically induced transparency (OMIT) is an interesting and important phenomenon. It was theoretically predicted by Agarwal and Huang \cite{Huang2010_041803} and experimentally observed in a microtoroid system \cite{Weis2010}, a superconducting circuit
cavity optomechanical system \cite{Teufel2011_471}, and a membrane-in-the-middle system \cite{Karuza2013}. More recently, the study of OMIT has attracted much attentions \cite{HY2015,Li2016SR,Shahidani2013,Chen2011,ZhangXY2018,Yan2015,Jia2015,Yan2014,ZhangH2018,Safavi-Naeini2011,LiuYX2013,Kronwald2013,Huang2011,Jing2015,Lu2017,Lu2018,Ma2014OL,Dong2013,Dong2015,Ma2014pra,Xiong2012,LiuYC2017,Xiong2018}. For instance, Huang studied OMIT in a quadratically coupled optomechanical systems where two-phonon processes occur \cite{Huang2011}. Jing et al. studied OMIT in a parity-time symmetric microcavity with a tunable gain-to-loss ratio \cite{Jing2015}. L$\ddot{\mathrm{u}}$ et al. studied OMIT in a spinning optomechanical system \cite{Lu2017}, and also studied OMIT at exceptional points \cite{Lu2018}. Ma et al. studied OMIT in the mechanical-mode splitting regime \cite{Ma2014OL}. Dong et al. studied the transient
phenomenon of OMIT \cite{Dong2013} and the 
Brillouin scattering induced transparency in a high-quality whispering-gallery-mode optical 
microresonator \cite{Dong2015}. Ma et al. studied tunable double OMIT in a hybrid optomechanical system with Coulomb coupling \cite{Ma2014pra}. Kronwald et al. studied OMIT in the nonlinear quantum regime \cite{Kronwald2013}. Xiong et al. studied OMIT in higher-order sidebands \cite{Xiong2012}, and the review articles on OMIT can be found in Refs. \cite{LiuYC2017,Xiong2018}.

The most prominent application of OMIT is light delay and storage \cite{Safavi-Naeini2011,Chen2011,LiuYC2017} due to the abnormal dispersion accompanied with the narrow transparency
window. Hence, having both a large depth and a small width at the transparency window is important for OMIT.  
Although increasing the power of the control field can lead to the increase of transparency depth, at the same time the width of the transparency window also increases. In addition, the ideal depth of the transparency window cannot be achieved due to the nonzero mechanical damping rate. These problems can be resolved if we consider a nonlinear effect in the response of the optomechanical system to the probe field.

In this paper, we mainly study OMIT in the simplest optomechanical model, described in Fig. (1), considering a nonlinear effect which was ignored in previous works. The Hamiltonian of the system is nonlinear and we can solve the nonlinear Heisenberg-Langevin equations using the perturbation method since the probe field is much weaker than the driving field. Note that if we linearize the nonlinear Heisenberg-Langevin equations following the usual linearization procedure \cite{Marquardt2007,Wilson-Rae2007,Vitali2007}, then the key nonlinear term will not exist in the response of the optomechanical system to the probe field. Considering the nonlinear term, we obtain the conditions for OMIT and find it has a strong impact on the absorptive and dispersive behavior of the optomechanical system to the probe field. First, the ideal depth of OMIT can be achieved easily even with nonzero mechanical damping rate, and there is only one suitable driving strength that can make the ideal OMIT occur. Secondly, the width of the transparency window depends only on three parameters of the system, and can become very narrow for small mechanical damping rate especially in unresolved sideband regime. And thirdly, if the driving strength continues to increase, the system will exhibit optomechanically induced gain, and the gain will become very large in unresolved sideband regime.

\section{System and Equations}

\begin{figure}[ptb]
	\includegraphics[width=0.45\textwidth]{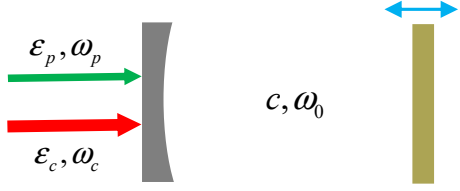}\caption{(Color online) Sketch of the standard optomechanical system consist of one mechanical resonator interacting with a cavity via radiation pressure effects. The cavity is driven by a coupling field with
	frequency $\omega_{c}$ (amplitude $\varepsilon_{c}$) and a weak probe field with frequency $\omega_{p}$ (amplitude $\varepsilon_{p}$).}%
	\label{Fig1}%
\end{figure}

We consider the standard optomechanical model in which a cavity is
coupled to a mechanical resonator (e.g. a moveable end-mirror) with frequency $\omega_{m}$ via radiation pressure effects (see Fig. 1). The cavity annihilation (creation) operator is denoted by $c$ ($c^{\dagger}$) with the
commutation relation $[c$, $c^{\dagger}]=1$. Momentum and position operators of the mechanical resonator with mass $m$ and damping rate $\gamma$ are
represented by $p$ and $q$, respectively. The mechanical resonator
makes small oscillations under the action of the radiation pressure force exerted by the
photons within the cavity. In turn, the mechanical displacement $q$ modifies the cavity resonance
frequency, represented by $\omega_{0}(q)$ which can be expanded to leading order in the displacement with $L$ the cavity length and $\omega_{0}$ the optical resonance frequency for $q=0$, i.e., $\omega_{0}(q)\approx\omega_{0}(1-\frac{q}{L})$ \cite{Aspelmeyer2014}. Hence, the interaction Hamiltonian between the cavity
and mechanical resonator can be described by $-\chi _{0}c^{\dagger}cq$ with $\chi _{0}=\hbar\omega _{0}/L$ being the optomechanical coupling constant.
The cavity is driven by a strong coupling field with
frequency $\omega_{c}$ (amplitude $\varepsilon_{c}$) and a weak probe field with frequency $\omega_{p}$ (amplitude $\varepsilon_{p}$). We
define $\wp_{c}$ and $\wp_{p}$ as the input powers of
relevant fields and $\kappa$ as the cavity decay rate, then the field amplitudes can be described as $\varepsilon_{c}=\sqrt{2\kappa\wp_{c}/(\hbar\omega_{c})}$ and $\varepsilon_{p}=\sqrt{2\kappa\wp_{p}/(\hbar\omega_{p})}$.  Thus, the Hamiltonian of the system can be written as
\begin{eqnarray}
H &=&\hbar\omega _{0}c^{\dagger }c+\frac{p^{2}}{2m}+\frac{1}{2}m\omega_{m}^{2}q^{2}
+i\hbar
\varepsilon_{c}(c^{\dagger}e^{-i\omega_{c}t}-ce^{i\omega_{c}t})\notag\\
&+&i\hbar(c^{\dagger}\varepsilon_{p}e^{-i\omega_{p}t}-c\varepsilon_{p}^{\ast}e^{i\omega_{p}t})-\chi_{0}c^{\dagger}cq.  
\end{eqnarray}%

In this paper, we deal with the mean response of the system to the probe field in the presence of the coupling field, hence we do not include quantum fluctuations. 
We use the factorization assumption $\langle qc\rangle=\langle q\rangle\langle
c\rangle$ and also
transform the cavity field to a rotating frame at the frequency $\omega_{c}$, the mean value equations are then given by
\begin{eqnarray}
\langle \dot{q}\rangle &=&\frac{\langle p\rangle }{m},\\
\langle\dot{p}\rangle &=&-m\omega_{m}^{2}\langle
q\rangle+\chi _{0}\langle c^{\dagger }\rangle \langle c\rangle-\gamma\langle p\rangle,\notag\\
\langle\dot{c}\rangle &=&-[\kappa +i(\omega_{0}-\omega_{c}-\chi_{0}\langle q\rangle/\hbar)]\langle c\rangle
+\varepsilon_{c}+\varepsilon_{p}e^{-i\delta t}.\notag
\end{eqnarray}
Here, $\delta=\omega_{p}-\omega_{c}$ is the detuning between probe field and coupling field.
Equations (2) are nonlinear, and therefore some of nonlinear effects will be omitted if we linearize Eq. (2) by the usual linearization method.
Considering that the probe field $\varepsilon_{p}$ is much weaker than the coupling field $\varepsilon_{c}$, we can solve Eq. (2) by the perturbation method attaining its steady-state solutions just to the first order in $\varepsilon_{p}$, i.e., $\langle s\rangle=s_{0}+\varepsilon_{p}e^{-i\delta t}s_{+}+\varepsilon^{\ast}_{p}e^{i\delta t}s_{-}$ $(s=q$, $p$, $c)$. The solutions, such as $c_{0}=\varepsilon_{c}/(\kappa+i\Delta)$ with the effective detuning $\Delta=\omega_{0}-\omega_{c}-\chi_{0}q_{0}/\hbar$, can be easily obtained. We will not list them one by one, because here we just care about the field with frequency $\omega_{p}$ in the output field. 

According the input-output relation \cite{Huang2010_041803,Walls}, the quadrature of the optical components with frequency $\omega_{p}$ in the output field
can be defined as $\varepsilon_{T}=2\kappa c_{+}$ \cite{Huang2010_041803}. The real part $\mathrm{Re}[\varepsilon_{T}]$ and imaginary part $\mathrm{Im}[\varepsilon_{T}]$ represent
the absorptive and dispersive behavior of the optomechanical system to the probe field, respectively. Because it is known that the coupling between the cavity and the resonator is strong at the near-resonant frequency, here we consider $\delta\sim\Delta\sim\omega_{m}$ and set $x=\delta-\omega_{m}$. After some calculations (see the detailed calculations in Appendix A), the result of $\varepsilon_{T}$ can be obtained as
\begin{eqnarray}
\varepsilon_{T}=\frac{2\kappa}{\kappa-ix+\frac{\beta}{\frac{\gamma}{2}-ix+\mathcal{N}}}
\end{eqnarray}
where 
\begin{eqnarray}
\mathcal{N}&=&-\frac{\beta}{\kappa-2i\omega_{m}},\\
\beta&=&\frac{\chi^{2}_{0}\varepsilon^{2}_{c}}{2m\omega_{m}\hbar(\kappa^{2}+\omega^{2}_{m})}.
\end{eqnarray}
The term $\mathcal{N}$ is the key term which will not exist in the subfraction of Eq. (3) if we adopt the usual linearization method to solve Eq. (2). Hence the origin of this term should be nonlinear effects and we call $\mathcal{N}$ the nonlinear term in the following.

With the nonlinear term, the conditions of ideal OMIT dip can be easily obtained. It can be obviously seen from Eq. (3) that the location of the pole in the subfraction of Eq. (3) can give the conditions. According to the location of the pole, setting $\frac{\gamma}{2}-ix+\mathcal{N}=0$, the conditions can be obtained as
\begin{eqnarray}
x&=&x_{o}\equiv-\frac{\gamma\omega_{m}}{\kappa},\\
\beta&=&\beta_{o}\equiv\frac{\gamma(\kappa^{2}+4\omega^{2}_{m})}{2\kappa}.
\end{eqnarray}
Equation (6) gives the concrete location $x_{o}$ where the ideal optomechanically induced transparency dip appears, and Equation (7) gives the suitable $\beta_{o}$ (corresponding to the suitable amplitude $\varepsilon_{c}$ of coupling field according to Eq. (5)) that can make the ideal OMIT dip occur even with nonzero mechanical damping rate $\gamma$. While if we ignore the nonlinear term $\mathcal{N}$, the real part $\mathrm{Re}[\varepsilon_{T}]=\frac{2\gamma\kappa}{2\beta+\gamma\kappa}$ at transparency window ($x=0$), which means that the ideal optomechanically induced transparency dip cannot appear because $\gamma\neq0$ and moreover it is necessary to increase coupling strength $\beta$ for large transparency depth.

It is worth pointing out that $\beta>\beta_{o}$ means that $\mathrm{Re}[\varepsilon_{T}]<0$, corresponding to gain which will not occur if we ignore the nonlinear term $\mathcal{N}$. Next, we first study the properties of OMIT with $\beta=\beta_{o}$ and then the properties of optomechanically induced gain with $\beta>\beta_{o}$. In this paper, we focus on the most studied regime of optomechanics where $\gamma\ll\omega_{m},$ $\kappa$, and set the mechanical quality factor as $Q=\omega_{m}/\gamma=10^{4}$ in the following.

\section{Optomechanically induced transparency}

From the above analysis, the nature of OMIT is determined by only three parameters, i.e., $\gamma$, $\omega_{m}$ and $\kappa$.
The width $\Gamma_{\mathrm{OMIT}}$ (full width at half maximum) of the transparency window is an important index in OMIT. According to Eq. (3) and the condition in Eq. (7), the analytical expression of the width $\Gamma_{\mathrm{OMIT}}$ (see the detailed calculations in Appendix B) can be obtained as
\begin{eqnarray}
\Gamma_{\mathrm{OMIT}}&=&\frac{\sqrt{\kappa^{4}+2\gamma\kappa(\kappa^{2}+\kappa\omega_{m}+4\omega^{2}_{m})}}{2\kappa}\notag\\
&+&\frac{\sqrt{\kappa^{4}+2\gamma\kappa(\kappa^{2}-\kappa\omega_{m}+4\omega^{2}_{m})}}{2\kappa}-\kappa
\end{eqnarray}
which can be simplified as
\begin{eqnarray}
\Gamma_{\mathrm{OMIT}}=\gamma(1+\frac{4\omega^{2}_{m}}{\kappa^{2}})
\end{eqnarray}
if $\gamma\omega^{2}_{m}\ll\kappa^{3}$. 
Equation (9) means the width $\Gamma_{\mathrm{OMIT}}$ can be very narrow especially in unresolved sideband regime. While if the nonlinear term $\mathcal{N}$ is ignored, the width $\Gamma_{\mathrm{OMIT}}\approx\frac{\gamma}{2}+\frac{\beta}{\kappa}$ \cite{Huang2010_041803} which means in this case the width will become very large due to the large $\beta$ needed to increase the depth of transparency.
We will discuss the properties of OMIT in the case of resolved sideband and unresolved sideband regime respectively in the following.

\subsection{Resolved sideband regime}

In the resolved sideband regime, i.e., $\kappa\ll\omega_{m}$, the width $\Gamma_{\mathrm{OMIT}}$ in Eq. (9) will become
\begin{eqnarray}
\Gamma_{\mathrm{OMIT}}&=&\frac{4\gamma\omega^{2}_{m}}{\kappa^{2}}.
\end{eqnarray}
It can be seen from Eq. (10) that the width $\Gamma_{\mathrm{OMIT}}$ is much larger than mechanical damping rate $\gamma$ in resolved sideband regime.
In Fig. (2), we plot the real part of $\varepsilon_{T}$ vs. the normalized frequency detuning $x/\gamma$ with $\beta=\beta_{o}$ according to Eq. (7) and with resolved sideband parameters $\omega_{m}=5\kappa$ which are similar to those in an optomechanical experiment on the observation of the normal-mode splitting \cite{Nature460}. 
According to Eq. (10), the width $\Gamma_{\mathrm{OMIT}}=100\gamma$ for $\omega_{m}=5\kappa$, which shows an excellent agreement with the numerical result in Fig. (2).

\begin{figure}[ptb]
	\includegraphics[width=0.45\textwidth]{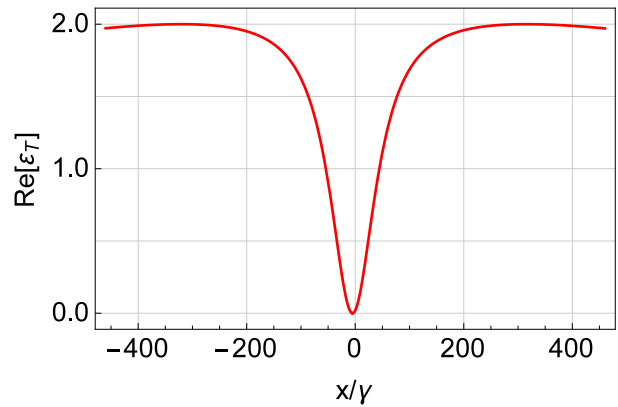}\caption{(Color online) The real part of $\varepsilon_{T}$ vs. normalized frequency detuning $x/\gamma$ with $\beta=\beta_{o}$ according to Eq. (7) and with resolved sideband parameters $\omega_{m}=5\kappa$.}%
	\label{Fig2}%
\end{figure}

\begin{figure}[ptb]
	\includegraphics[width=0.46\textwidth]{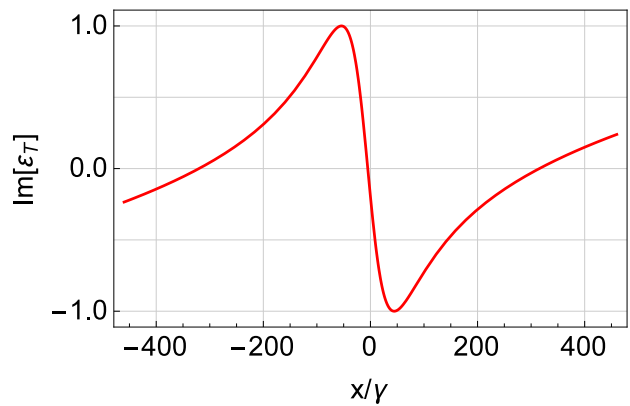}\caption{(Color online) The imaginary part of $\varepsilon_{T}$ vs. normalized frequency detuning $x/\gamma$ with the same parameters in Fig. (2).}%
	\label{Fig3}%
\end{figure}

The dispersive behavior, represented by $\mathrm{Im}[\varepsilon_{T}]$, is related to slow light effects of the optomechanical system to the probe field. 
In Fig. (3), we plot the imaginary part of $\varepsilon_{T}$ vs. the normalized frequency detuning $x/\gamma$ with $\beta_{o}$ according to Eq. (7) and with the same parameters in Fig. (2).
It can be seen from Fig. (3) that the steepest dispersion occurs at the point $x_{o}$ where the OMIT dip appears. The negative maximum value of the dispersion curve slope can be obtained (see the detailed calculations in Appendix C) as
\begin{eqnarray}
\mathrm{K}_{\mathrm{max}}&=&-\frac{4\kappa^{2}}{\gamma(\kappa^{2}+4\omega^{2}_{m})}.
\end{eqnarray}
Note that Eq. (11) is true for both resolved sideband and unresolved sideband regime.
From Eq. (9) and (11), we have
\begin{eqnarray}
\mathrm{K}_{\mathrm{max}}\times\Gamma_{\mathrm{OMIT}}&=&-4
\end{eqnarray}
which means that the narrower the width $\Gamma_{\mathrm{OMIT}}$ is, the steeper the dispersion curve becomes.

\subsection{Unresolved sideband regime}

Compared with the case of resolved sideband regime, according to Eq. (9), the width $\Gamma_{\mathrm{OMIT}}$ will become more narrower in unresolved sideband regime. 
In Fig. (4), we plot the real part of $\varepsilon_{T}$ vs. the normalized frequency detuning $x/\gamma$ with $\beta=\beta_{o}$ according to Eq. (7) and with unresolved sideband parameters
$\kappa=2\omega_{m}$ (blue dashed line) and $\kappa=5\omega_{m}$ (red solid line).
According to Eq. (9), the width $\Gamma_{\mathrm{OMIT}}=2\gamma$ for $\kappa=2\omega_{m}$ and $\Gamma_{\mathrm{OMIT}}=1.16\gamma$ for $\kappa=5\omega_{m}$. These results are consistent with the numerical results in Fig. (4).

The imaginary part of $\varepsilon_{T}$ vs. normalized frequency detuning $x/\gamma$ is plotted in Fig. (5) with the same parameters in Fig. (4).  From Fig. (5), it can be clearly seen that the dispersion curve becomes steeper with larger ratio of $\kappa/\omega_{m}$ and the negative maximum value of the dispersion curve slope is still given by Eq. (11).

\begin{figure}[ptb]
	\includegraphics[width=0.45\textwidth]{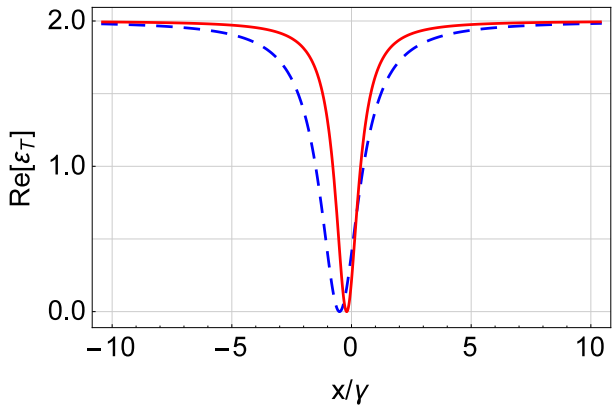}\caption{(Color online) The real part of $\varepsilon_{T}$ vs. normalized frequency detuning $x/\gamma$ with $\beta=\beta_{o}$ according to Eq. (7) and with unresolved sideband parameters $\kappa=2\omega_{m}$ (blue dashed line), $\kappa=5\omega_{m}$ (red solid line).}%
	\label{Fig4}%
\end{figure}

\begin{figure}[ptb]
	\includegraphics[width=0.46\textwidth]{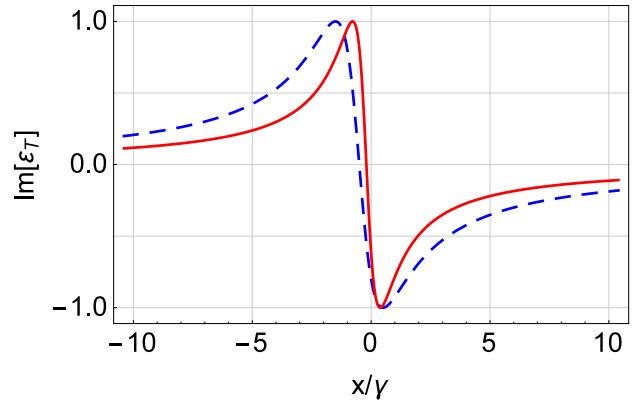}\caption{(Color online) The imaginary part of $\varepsilon_{T}$ vs. normalized frequency detuning $x/\gamma$. The parameters are the same in Fig. (4).}%
	\label{Fig5}%
\end{figure}

\section{Optomechanically induced gain}

Now we discuss the properties of optomechanically induced gain ($\mathrm{Re}[\varepsilon_{T}]<0$) which will occur when $\beta>\beta_{o}$. According to Eq. (3), we can numerically obtain all the properties of gain including the maximum gain value and the point $x$ where the gain takes the maximum. While if $\beta\gtrsim\beta_{o}$, the point where the gain takes the maximum can be given by
\begin{eqnarray}
x=x_{\mathrm{g}}\equiv-\frac{2\beta\omega_{m}}{\kappa^{2}+4\omega^{2}_{m}}
\end{eqnarray}
according to the same condition ($\mathrm{Im}[\frac{\gamma}{2}-ix+\mathcal{N}]=0$) as that we use to obtain the location $x_{o}$ in Eq. (6). For simplicity, we just discuss the properties of gain with $x=x_{\mathrm{g}}$ according to Eq. (13) in the following.

 \begin{figure}[ptb]
 	\includegraphics[width=0.46\textwidth]{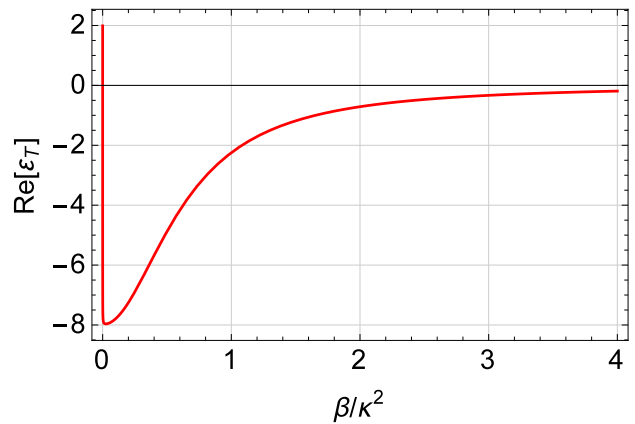}\caption{(Color online)  The real part of $\varepsilon_{T}$ vs. normalized drive strength $\beta/\kappa^{2}$ with $x=x_{g}$ according to Eq. (13) and with parameter $\kappa=4\omega_{m}$.}%
 	\label{Fig6}%
 \end{figure}
 
 \begin{figure}[ptb]
 	\includegraphics[width=0.46\textwidth]{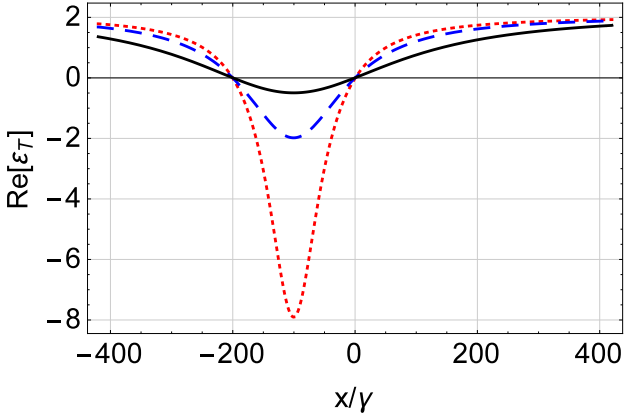}\caption{(Color online) the real part of $\varepsilon_{T}$ vs. the normalized frequency detuning $x/\gamma$ with $\beta=\beta_{g}$ according to Eq. (14) and with $\kappa=\omega_{m}$ (black solid line), $\kappa=2\omega_{m}$ (blue dashed line), and $\kappa=4\omega_{m}$ (red dotted line).}%
 	\label{Fig7}%
 \end{figure}

In fact, the gain does not always increase with the increase of $\beta$. The reason is that the negative value of $\mathrm{Re}[\varepsilon_{T}]$ approaches zero as $\beta\rightarrow\infty$ according to Eq. (3).
In Fig. (6), we plot $\mathrm{Re}[\varepsilon_{T}]$ vs. $\beta/\kappa^{2}$ with $x=x_{g}$ according to Eq. (13) and with $\kappa=4\omega_{m}$. It can be clearly seen from Fig. (6) that there exist an optimum value $\beta$, defined as $\beta_{\mathrm{g}}$, that makes $\mathrm{Re}[\varepsilon_{T}]$ take the maximum negative value $\mathcal{G}_{\mathrm{max}}$.

With Eq. (3) and Eq. (13), the numerical result of the optimum value $\beta_{\mathrm{g}}$ and the corresponding maximum negative value $\mathcal{G}_{\mathrm{max}}$ can be easily found out. However, the approximate analytic expression of $\beta_{\mathrm{g}}$ and $\mathcal{G}_{\mathrm{max}}$ can be obtained as
\begin{eqnarray}
\beta_{\mathrm{g}}&=&\frac{(\kappa^{2}+4\omega^{2}_{m})\sqrt{\gamma/\omega_{m}}}{2},\\
\mathcal{G}_{\mathrm{max}}&=&-\frac{2\kappa^{2}}{4\omega^{2}_{m}+\kappa\sqrt{\gamma\omega_{m}}}
\end{eqnarray}
if $\kappa\gg\sqrt{\gamma\omega_{m}}$. 

According to Eq. (15), we have the maximum negative value $\mathcal{G}_{\mathrm{max}}\approx-\kappa^{2}/2\omega_{m}^{2}\ll1$ in resolved sideband regime. However, the maximum negative value $\mathcal{G}_{\mathrm{max}}$ can become very large in unresolved sideband regime. In Fig. (7), we plot the real part of $\varepsilon_{T}$ vs. the normalized frequency detuning $x/\gamma$ with $\beta=\beta_{g}$ according to Eq. (14) and with $\kappa=\omega_{m}$ (black solid line), $\kappa=2\omega_{m}$ (blue dashed line), and $\kappa=4\omega_{m}$ (red dotted line).
According to Eq. (15), the maximum negative value
$\mathcal{G}_{\mathrm{max}}=-0.499$ for $\kappa=\omega_{m}$, $\mathcal{G}_{\mathrm{max}}=-1.990$ for $\kappa=2\omega_{m}$, and $\mathcal{G}_{\mathrm{max}}=-7.921$ for $\kappa=4\omega_{m}$. These results are consistent with the numerical results in Fig. (7).

\section{Conclusions}

In summary, we have theoretically studied the properties of optomechanically induced transparency in the simplest optomechanical system (consisting of one cavity and one mechanical resonator) with nonlinear effect that was ignored in previous works. We attain the conditions where the system can exhibit perfect optomechanically induced transparency, and obtain the expression of the width of optomechanically induced transparency dip. From these crucial expressions, we can draw three important conclusions: (1) there exist only one suitable driving strength that can make the ideal optomechanically induced transparency dip occur, and the properties of optomechanically induced transparency are determined by only three system parameters ($\gamma$, $\kappa$ and $\omega_{m}$); (2) the width of optomechanically induced transparency dip can become very narrow in unresolved sideband regime, and the product of the width and the dispersion slope at the transparency window is a constant; (3) the maximum value of optomechanically induced gain is very small in resolved sideband regime, while it can become very large in unresolved sideband regime. We believe these results can be used to control optical transmission in quantum information processing.

\appendix{}

\section{Derivation of $\varepsilon_{T}$}

To solve Eqs. (2), we substitute the formal solution $\langle s\rangle=s_{0}+\varepsilon_{p}e^{-i\delta t}s_{+}+\varepsilon^{\ast}_{p}e^{i\delta t}s_{-}$ $(s=q$, $p$, $c)$ into Eqs. (2) and keep its steady-state solutions only to the first order in $\varepsilon_{p}$. It is straight forward to obtain
\begin{eqnarray}
q_{0}&=&\frac{\chi_{0}\left\vert c_{0}\right\vert^{2}}{m\omega_{m}^{2}},\\
q_{+}&=&\frac{\chi_{0}(c_{0}c_{-}^{\ast}+c_{0}^{\ast}c_{+})}{m(\omega_{m}^{2}-i\delta\gamma-\delta^{2})},\\
q_{-}&=&\frac{\chi_{0}(c_{0}c_{+}^{\ast}+c_{0}^{\ast}c_{-})}{m(\omega_{m}^{2}+i\delta\gamma-\delta^{2})},\\
c_{0}&=&\frac{\varepsilon_{c}}{\kappa+i\Delta},\\
c_{+}&=&\frac{ic_{0}q_{+}\chi_{0}/\hbar+1}{\kappa+i(\Delta-\delta)},\\
c_{-}&=&\frac{ic_{0}q_{-}\chi_{0}/\hbar}{\kappa+i(\Delta+\delta)},
\end{eqnarray}
with $\Delta=\omega_{0}-\omega_{c}-q_{0}\chi_{0}/\hbar$.

From Eqs. (A2) and (A3), we have $q_{+}=q_{-}^{\ast}$, and
according to Eqs. (A2) and (A6), it can be obtained that 
\begin{eqnarray}
c_{0}c_{-}^{\ast}=\frac{\mathrm{M}}{1-\mathrm{M}}c_{0}^{\ast}c_{+}
\end{eqnarray}
with
\begin{eqnarray}
\mathrm{M}=\frac{-i\left\vert c_{0}\right\vert^{2}\chi_{0}^{2}}{m\hbar(\omega_{m}^{2}-i\delta\gamma-\delta^{2})(\kappa-i(\Delta+\delta))}.
\end{eqnarray}
By substituting Eq. (A7) into Eq. (A2) and according to Eq. (A5), we obtain the expression of $c_{+}$ as
\begin{eqnarray}
c_{+}=\frac{1}{\kappa-i(\delta-\Delta)+\frac{\beta}{\frac{\delta^{2}-\omega_{m}^{2}+i\delta\gamma}{2i\omega_{m}}-\frac{\beta}{\kappa-i(\Delta+\delta)}}}
\end{eqnarray}
with
\begin{eqnarray}
\beta=\frac{\chi_{0}^{2}\left\vert c_{0}\right\vert^{2}}{2m\hbar\omega_{m}}.
\end{eqnarray}
Due to the condition of near-resonant frequency, i.e., $\delta\sim\Delta\sim\omega_{m}$, we have $\delta^{2}-\omega_{m}^{2}\sim2\omega_{m}(\delta-\omega_{m})$ and $\delta+\Delta\sim2\omega_{m}$.
If we set $x=\delta-\omega_{m}$, Eq. (A9) can be simplified as
\begin{eqnarray}
c_{+}=\frac{1}{\kappa-ix+\frac{\beta}{\frac{\gamma}{2}-ix-\frac{\beta}{\kappa-2i\omega_{m}}}}.
\end{eqnarray}
Finally, we obtain the quadrature $\varepsilon_{T}$ by multipling $c_{+}$ by $2\kappa$, see Eq. (3). Besides, substituting Eq. (A4) into Eq. (A10), we obtain Eq. (5).

\section{Derivation of width $\Gamma_{\mathrm{OMIT}}$}

When the OMIT occurs, the conditions in Eqs. (6) and (7) must be satisfied.
By substituting Eq. (7) into Eq. (3) and making a translation transformation $x=y+x_{o}$ ($x_{o}=-\frac{\gamma\omega_{m}}{\kappa}$ according to Eq. (6)), Eq. (3) can be shown as
\begin{eqnarray}
\varepsilon_{T}=\frac{2\kappa}{\kappa+i(\frac{\gamma(\kappa^{2}+4\omega_{m}^{2})}{2y\kappa}+\frac{\gamma\omega_{m}}{\kappa}-y)}.
\end{eqnarray}
It can be seen from Eq. (B1) that $\mathrm{Re}[\varepsilon_{T}]=1$ (half maximum) when
\begin{eqnarray} \frac{\gamma(\kappa^{2}+4\omega_{m}^{2})}{2y\kappa}+\frac{\gamma\omega_{m}}{\kappa}-y=\pm\kappa.
\end{eqnarray}

From Eq. (B2), we can obtain the two roots $y_{1}$ and $y_{2}$ between which the difference gives the full width at half maximum, i.e., $\Gamma_{\mathrm{OMIT}}=y_{1}-y_{2}$. The concrete expressions of the two roots are
\begin{eqnarray} 
y_{1}=\frac{\gamma\omega_{m}-\kappa^{2}+\sqrt{(\gamma\omega_{m}-\kappa^{2})^{2}+2\gamma\kappa(\kappa^{2}+4\omega_{m}^{2})}}{2\kappa},\quad\\
y_{2}=\frac{\gamma\omega_{m}+\kappa^{2}-\sqrt{(\gamma\omega_{m}+\kappa^{2})^{2}+2\gamma\kappa(\kappa^{2}+4\omega_{m}^{2})}}{2\kappa}.\quad 
\end{eqnarray}
Due to $\gamma\ll\kappa$, we can safely ignore the quadratic term $\gamma^{2}\omega_{m}^{2}$ in Eqs. (B3) and (B4), and then we obtain the full width $\Gamma_{\mathrm{OMIT}}$, see Eq. (8).

\section{Dispersion curve slope  $\mathrm{K}_{\mathrm{max}}$}

The imaginary part $\mathrm{Im}[\varepsilon_{T}]$ represents
the dispersive behavior of the optomechanical system to the probe field.
According to Eq. (B1), we can easily obtain the imaginary part $\mathrm{Im}[\varepsilon_{T}]$ as
\begin{eqnarray} 
\mathrm{Im}[\varepsilon_{T}]=\frac{4y\kappa^{2}(2y^{2}\kappa-2y\gamma\omega_{m}-\eta)}{4y^{4}\kappa^{2}-8y^{3}\gamma\kappa\omega_{m}+y^{2}\xi+4y\gamma\omega_{m}\eta+\eta^{2}},\quad
\end{eqnarray}
with
\begin{eqnarray} 
\eta&=&\gamma(\kappa^{2}+4\omega_{m}^{2}),\\
\xi&=&4(\kappa^{4}+\gamma^{2}\omega_{m}^{2}-\gamma\kappa^{3}-4\gamma\kappa\omega_{m}^{2}).
\end{eqnarray}

According to Eq. (C1), the dispersion curve slope $\mathrm{K}$ $(=\frac{\partial\mathrm{Im}[\varepsilon_{T}]}{\partial y})$ can be obtained as
\begin{eqnarray} 
\mathrm{K}=\frac{4y^{2}\kappa^{2}(2y^{2}\kappa+\eta)(\xi-4y^{2}\kappa^{2}+8y\gamma\kappa\omega_{m}-8\gamma^{2}\omega_{m}^{2})}{(\eta^{2}+4y\gamma\eta\omega_{m}+4y^{4}\kappa^{2}+y^{2}\xi-8y^{3}\gamma\kappa\omega_{m})^{2}}\nonumber\\
-\frac{4\eta\kappa^{2}(\eta+2y^{2}\kappa)(\eta+4y\gamma\omega_{m}-8y^{2}\kappa)}{(\eta^{2}+4y\gamma\eta\omega_{m}+4y^{4}\kappa^{2}+y^{2}\xi-8y^{3}\gamma\kappa\omega_{m})^{2}}.\;\qquad
\end{eqnarray}
The dispersion curve slope will take the maximum $\mathrm{K}_{\mathrm{max}}$ at the transparency window where $x=-\frac{\gamma\omega_{m}}{\kappa}$ ($y=0$).
From Eq. (C4), setting $y=0$, we can obtain the expression of $\mathrm{K}_{\mathrm{max}}$, see Eq. (11).

\end{document}